\documentclass{snmp2003}

\def\R{{\mathbb R}}
\def\S{{\mathbb S}}

\begin{document}

\FirstPageHeading{Bronnikov}

\ShortArticleName{Self-Gravitating Global Monopole and Nonsingular Cosmology}

\ArticleName{Self-Gravitating Global Monopole\\ and Nonsingular Cosmology}

\Author{K.A. BRONNIKOV~$^\dag$ and B.E. MEIEROVICH~$^\ddag$}
\AuthorNameForHeading{K.A. Bronnikov and B.E. Meierovich}
\AuthorNameForContents{Bronnikov K.A. and Meierovich B.E.}
\ArticleNameForContents
     {Self-Gravitating Global Monopole and Nonsingular Cosmology}

\Address{$^\dag$~VNIIMS, 3-1 M. Ulyanovoy St., Moscow 117313, Russia;\\
          Institute of Gravitation and Cosmology, PFUR,
          6 Miklukho-Maklaya St., Moscow 117198, Russia}
\EmailD{kb@rgs.mccme.ru}
\Address{$^\ddag$~Kapitza Institute for Physical
          Problems, 2 Kosygina St., Moscow 117334, Russia}
\EmailD{meierovich@yahoo.com}

\Abstract{We review some recent results concerning the properties of a
    spherically symmetric global monopole in $(D=d+2)$-dimensional general
    relativity. Some common features of monopole solutions are found
    independently of the choice of the symmetry-breaking potential. Thus,
    the solutions show six types of qualitative behavior and can contain at
    most one simple horizon. For the standard Mexican hat (Higgs) potential,
    we analytically find the $D$-dependent range of the parameter $\gamma$
    (characterizing the gravitational field strength) in which there exist
    globally regular solutions with a monotonically growing Higgs field,
    containing a horizon and a Kantowski-Sachs (KS) cosmology outside it,
    with the topology of spatial sections $\R\times \S^d$. Their
    cosmological properties favor the idea that the standard Big Bang might
    be replaced with a nonsingular static core and a horizon appearing as a
    result of some symmetry-breaking phase transition on the Planck energy
    scale. We have also found families of new solutions with an oscillating
    Higgs field, parametrized by the number of its knots. All such solutions
    describe space-times of finite size, possessing a regular center, a
    horizon and a singularity beyond it.}

    According to the Standard cosmological model, the Universe has been
    expanding and cooling from a split second after the Big Bang to the
    present time and remained uniform and isotropic on the large in doing so.
    In the process of its evolution, the Universe has experienced a chain
    of phase transitions with spontaneous symmetry breaking, including
    Grand Unification and electroweak phase transitions, formation of
    neutrons and protons from quarks, recombination, and so forth. Regions
    with spontaneously broken symmetry which are more than the correlation
    length apart, are statistically independent. At interfaces between
    these regions, the so-called topological defects necessarily arise.

    A systematic exposition of the potential role of topological defects
    in our Universe has been provided by Vilenkin and Shellard
    \cite{Bronnikov:Vilenkin and Shellard}; see there also the necessary
    references to the previous work. The particular types of defects:
    domain walls, strings, monopoles, or textures are determined by the
    topological properties of vacuum. If the vacuum manifold after the
    breakdown is not shrinkable to a point, then solutions of Polyakov-'t
    Hooft monopole type appear in quantum field theory.

    Spontaneous symmetry breaking is well known to play a fundamental role
    in modern attempts to construct particle theories. In this context,
    one mostly deals with internal symmetries rather than those associated
    with space-time transformations: examples are the Grand Unification
    symmetry, the electroweak and isotopic symmetries and supersymmetry,
    whose transformations mix bosons and fermions. Topological defects,
    caused by spontaneous breaking of internal symmetries independent of
    space-time coordinates, are called global.

    A fundamental property of global symmetry violation is the Goldstone
    degree of freedom. In the monopole case, the term related to the
    Goldstone boson in the energy-momentum tensor decreases rather slowly
    away from the center. As a result, the total energy of a global
    monopole grows linearly with distance, in other words, diverges.
    Without gravity such a divergence is a general property of
    spontaneously broken global symmetries. The self-gravity of a global
    monopole, if not entirely removes this difficulty, allows one to
    consider it from a new standpoint.

    We have performed a general study of the properties of static,
    spherically symmetric global monopoles in general relativity
    \cite{Bronnikov:BM-02}. Most of the results were extended
    \cite{Bronnikov:BM-03} to spherically symmetric configurations of
    arbitrary dimension with the topology $\R \times \R_+ \times \S^d$, but
    in this presentation we shall for simplicity mostly adhere to $D=4$
    ($d=2$). In this case, the Lagrangian is taken in the form
\begin{equation}                                         \label{Bronnikov:L}
        L=\frac{1}{2}\partial _{\mu }\phi
        ^{a}\partial^{\mu }\phi^{a} - V (\phi) + \frac{R}{16\pi G},
\end{equation}
    where $R$ is the scalar curvature, $G$ is the gravitational constant,
    $\phi^a$, $a=1,2,3$ is a scalar field multiplet, and  $V (\phi) $ is a
    symmetry-breaking potential depending on $\phi =\pm \sqrt
    {\phi^a\phi^a}$. We assume the static, spherically symmetric metric
\begin{equation}                                        \label{Bronnikov:ds}
        ds^2 = A(\rho) dt^2 - \frac{d\rho^2}{A(\rho)}
                    -r^2(\rho) d\Omega^2
\end{equation}
    ($d\Omega^2 \equiv d\theta^2 + \sin^2\theta\, d\varphi^2$)
    and a ``hedgehog'' scalar field configuration:
\begin{equation}                                       \label{Bronnikov:hog}
        \phi^a = \phi(u) n^a, \qquad\
           n^a = \left\{\sin\theta \cos\varphi,\quad
                        \sin\theta \sin\varphi,\quad
                        \cos\theta               \right\}.
\end{equation}
    Without loss of generality we take $\rho\geq 0$ and attribute $\rho=0$
    to a regular center.

    Our approach was different from most of previous studies which had
    used the boundary condition that $V =0$ at spatial infinity. We did
    not even require the existence of a spatial asymptotic.
    Instead, we required regularity at the center and tried to observe the
    properties of the whole set of global monopole solutions. Also, instead
    of dealing only with a particular form of the symmetry breaking
    potential (usually the Mexican hat potential), we found some general
    features of solutions valid independently of the particular shape of
    $V(\phi)$.  The main results are as follows.

    The Einstein equations lead to $r''\leq 0$. Since at a regular center
    $r' >0$, this leaves three possibilities for the function $r(\rho)$:

\begin{description}
\item[(a)]
    monotonic growth with a decreasing slope, but $r\to \infty$ as
    $\rho\to\infty$,

\item[(b)]
    monotonic growth with $r\to r_{\max} <\infty$ as $\rho\to\infty$, and

\item[(c)]
    growth up to $r_{\max}$ at some $\rho_1 < \infty$ and further
    decrease, reaching $r=0$ at some finite $\rho_2 > \rho_1$.
\end{description}

    The other metric function, $A(\rho)$, determines the causal structure of
    space-time: zeros of any order of $A(\rho)$ correspond to Killing
    horizon of the same order.

\begin{proposition}
    Under the assumption that $\phi^2 < 1/(8\pi G)$ in the whole space, our
    system with a regular center can have either no horizon, or one simple
    horizon, and in the latter case its global structure is the same as that
    of de Sitter space-time.
\end{proposition}

    For nonnegative $V(\phi)$ there is no restriction on the magnitude of
    $\phi$.

\begin{proposition}
    If\ $V(\phi)\geq 0$, our system with a regular center can have either no
    horizon, or one simple horizon, and in the latter case its global
    structure is the same as that of de Sitter space-time.
\end{proposition}

    Outside the horizon, in the so-called T-region, $\rho$ becomes a
    temporal coordinate, and the geometry corresponds to homogeneous
    anisotropic cosmological models of Kantowski-Sachs (KS) type, where
    spatial sections have the topology $\R \times \S^2$.

    Thus, depending on the behavior of $r(\rho)$ (items a--c) and $A(\rho)$
    (with or without a horizon), all possible solutions can be divided into
    six qualitatively different classes. There are two more general results
    valid for any nonnegative potentials.

\begin{proposition}
    If\ $V(\phi)\geq 0$, the second center $r=0$, if any, is singular.
\end{proposition}

\begin{proposition}
    If\ $V(\phi)\geq 0$ and the solution is asymptotically flat, the mass
    $M$ of the global monopole is negative.
\end{proposition}

    In Proposition 4, asymptotic flatness is understood up to the solid angle
    deficit $\Delta < 1$: at large $\rho$ we have $r(\rho) \approx \rho$ and
\begin{equation}
    ds^2=\biggl(\frac{1-\Delta}{\alpha^2}
                    		- \frac{2GM}{r}\biggr) dt^2
       -\biggl(\frac{1-\Delta}{\alpha^2}
                -\frac{2GM}{r}\biggr)^{-1} dr^2
                    - r^2 d\Omega^2,       	   \label{Bronnikov:qSch}
\end{equation}
    where $\alpha$ is a model-dependent constant. To our knowledge,
    Proposition 4 had been obtained previously only numerically for the
    particular potential (\ref{Bronnikov:hat}) (see below).

    Thus, even before studying particular solutions, we have a more or less
    complete knowledge of what can be expected from such global monopole
    systems.

    We further considered \cite{Bronnikov:BM-02} the most frequently
    used Mexican hat potential
\begin{equation}
       V(\phi) = \frac 14 \lambda (\phi^a \phi^a - \eta^2)^2
               = \frac 14 \eta^4 \lambda (f^2 - 1)^2,  \label{Bronnikov:hat}
\end{equation}
    where $\eta > 0$ characterizes the energy of symmetry breaking,
    $\lambda$ is a dimensionless coupling constant and $f(u)= \phi(u)/\eta$
    is the normalized field magnitude playing the role of an order
    parameter. The model has a global $SO(3)$ symmetry, which can be
    spontaneously broken to $SO(2)$ due to the potential wells ($V=0$) at
    $f=\pm 1$.

    Our analytical and numerical study for the potential (\ref{Bronnikov:hat}) has
    confirmed previous results of other authors concerning configurations
    with a monotonically growing field magnitude $f$.

    The solution properties are basically governed by the values of the
    single dimensionless parameter
\begin{equation}
        \gamma =8\pi G\eta ^{2},                    \label{Bronnikov:gamma}
\end{equation}
    characterizing the gravitational field strength. Thus, for $\gamma < 1$
    the solutions have a spatial asymptotic with the metric
    (\ref{Bronnikov:qSch}).  For $1 < \gamma < 3$, the so-called
    supermassive global monopole, the solutions contain a cosmological
    horizon and a KS model outside it.

    We have obtained analytically the upper limit $\gamma_0=3$ for the
    existence of static monopole solutions, previously found numerically by
    Liebling \cite{Bronnikov:Liebling}. To do so, we have used the fact
    that near a critical value of $\gamma$ the field magnitude is small
    everywhere inside the horizon, making it possible to formulate a
    well-posed eigenvalue problem for the field $f$ against the background
    of the de Sitter metric (which solves the Einstein equations in case
    $f\equiv 0$).  The linear equation for $f$ has the form
\begin{equation}
    \frac{d}{dx} \biggl[ x^2 \biggl( 1-\frac{x^2}{x_h^2}
      \biggr)\frac{df}{dx} \biggr] -  (2-x^{2}) f=0, \label{Bronnikov:Eq-fr}
\end{equation}
    where $x$ is a dimensionless variable proportional to $r$ and
    $x_h = \sqrt{12/\gamma }$ is the value of $x$
    at the horizon. The boundary conditions are
\begin{equation}
    f\Big|_{x=0} =0, \qquad |f(x_h)| < \infty.  \label{Bronnikov:bound_lin}
\end{equation}
    Nontrivial solutions of (\ref{Bronnikov:Eq-fr}) with these boundary
    conditions exist for the sequence of eigenvalues
\begin{equation}
    \gamma_n = \frac{3}{(n+1/2)(n+2)}, \qquad  \label{Bronnikov:gam_n}
    n=0,1,2,...,
\end{equation}
    where $n$ is the number of nodes of the corresponding eigenfunctions
    $f_n (x)$. The eigenvalue $\gamma_0 = 3$ is the sought-for critical value
    of $\gamma$ for monotonically growing $f$. In case $\gamma > 3$ static
    monopole solutions are absent.

    The solutions with $n \geq 1$ form new families, which we had
    not met in the existing literature. In these solutions, existing for
    $\gamma < \gamma_n$, the field function $f$ (which has no reason to be
    small when $\gamma$ is far from $\gamma_n$), changes its sign $n$
    times.  All such solutions turn out to have a singularity ($f\to
    \infty$, $r\to 0$) at some finite value of $\rho$ beyond the horizon.

    The solutions with a static nonsingular monopole core and a KS
    cosmological model in a T-region ($A(\rho) < 0$) outside the horizon are
    of particular interest. Changing the notations, $t\to y \in \R$, and
    introducing the proper time of a comoving observer
    $\tau = \int d\rho/\sqrt{|A(\rho)|}$, we can rewrite the metric as
\begin{equation}
       ds^2 = d\tau^2 - |A(\tau)| dy^2 - r^2(\tau) d\Omega^2
\end{equation}
    The model expands in one of the directions (along the $y$ axis) from
    zero at the horizon (say, $\tau =0$) to finite values at large $\tau$ in
    a process which, on its early stage, resembles inflation. In two other
    directions, corresponding to $\S^2$, the model expands from a finite
    size and finite expansion rate at $\tau=0$ to a linear regime at large
    $\tau$. Like other regular models with the de Sitter causal structure,
    i.e., a static core and expansion beyond a horizon, these models
    drastically differ from standard Big Bang models in that the expansion
    starts from a nonsingular surface, and cosmological comoving observers
    can receive information in the form of particles and light quanta from
    the static region, situated in the absolute past with respect to them.
    Moreover, in our case the static core is nonsingular, and it is thus an
    example of an entirely nonsingular cosmology in the spirit of the views
    advocated by Gliner and Dymnikova (\cite{Bronnikov:Gliner,
    Bronnikov:Dymnikova}; see also references therein).

    The nonzero symmetry-breaking potential plays the role of a
    time-dependent cosmological constant, a kind of hidden vacuum matter.
    For an observer in the T-region the potential decreases with time, and
    the hidden vacuum matter gradually disappears.

    The present simple model cannot be directly applied to our Universe (in
    particular, due to lack of isotropization), it can at most pretend to
    describe the earliest, near-Planckian stage in an approximate,
    classical manner. It nevertheless may be considered as an argument in
    favor of the idea that the standard Big Bang might be replaced with a
    nonsingular static core and a horizon appearing as a result of some
    symmetry-breaking phase transition on the Planck energy scale.

    As another cosmological application of the global monopole, one should
    mention the concept of topological inflation, related to the existence
    of a de Sitter core of the monopole, which can inflate due to its
    instability \cite{Bronnikov:topoinfl}

    In Ref.\,\cite{Bronnikov:BM-03} we have extended the above consideration
    to $(D = d+2)$-dimensional general relativity, with the space-time
    topology $\R \times \R_+ \times \S^d$. The qualitative features of the
    solutions are mostly preserved, in particular, there are the same six
    types of behavior, and Propositions 1--4 still hold. Outside the horizon
    (if any), the metric again corresponds to a Kantowski-Sachs type
    cosmology, now with the topology of spatial sections $\R\times \S^d$.

    For the Mexican hat potential (\ref{Bronnikov:hat}), the strength
    parameter is defined as $\gamma = \kappa^2 \eta^2$ where $\kappa^2$ is
    the gravitational constant of $D$-dimensional theory and $\eta$ is the
    symmetry breaking characteristic from (\ref{Bronnikov:hat}). There are
    again two critical values of $\gamma$ for each $D$: for monotonic
    $f(\rho)$, solutions with static spatial infinity exist with $\gamma <
    d-1$, while solutions with a horizon and an infinitely expanding KS
    exterior correspond to $d-1 < \gamma < 2d(d+1)/(d+2)$. For solutions
    where $f(\rho)$ changes its sign $n$ times, instead of
    (\ref{Bronnikov:gam_n}), the critical values of $\gamma$ are
\begin{equation}
    \gamma_n = \frac{2d(d+1) }{(2n+1)(2n+d+2)}, \label{Bronnikov:ga_n}
\end{equation}
    which reduces to (\ref{Bronnikov:gam_n}) when $d=2$.

    In the important case when the horizon is far from the monopole core,
    the temporal evolution of the KS metric is described analytically. The
    Kantowski-Sachs space-time contains a subspace with a closed
    Friedmann-Robertson-Walker metric. Our estimates show that the
    5-dimensional global monopole model is in principle able to give
    plausible cosmological parameters. However, within our macroscopic
    theory without specifying the physical nature of vacuum we cannot
    unequivocally explain why the fourth spatial dimension (the one that
    played the role of time in the static region) is not observable.
    Quantitative estimates certainly require a more complete model including
    further phase transitions, one of which should explain the
    unobservable nature of the extra dimension.

\LastPageEnding
\end{document}